\documentclass{article}
\usepackage[numbers]{natbib}
\usepackage{float}
\pdfoutput=1
\usepackage{arxiv}
\usepackage{subcaption}
\usepackage{sidecap}

\usepackage{indentfirst}
\usepackage[utf8]{inputenc} 
\usepackage[T1]{fontenc}    
\usepackage{hyperref}     
\usepackage{url}         
\usepackage{booktabs}    
\usepackage{amsfonts}   
\usepackage{nicefrac}       
\usepackage{microtype}    
\usepackage{graphicx}
\usepackage{doi}

\graphicspath{ {./images/} }

\title{Predicting Post-Concussion Syndrome Outcomes With Machine Learning}

\author{{\hspace{1mm}Minhong Kim} \\\\
	Gyeonggi Suwon International School \\\\\\
}

\renewcommand{\shorttitle}{\emph{Predicting Post-Concussion Syndrome Outcomes}}

\begin{document}
\maketitle

\begin{abstract}
In this paper, machine learning models are used to predict outcomes for patients with persistent post-concussion syndrome (PCS). Patients had sustained a concussion at an average of two to three months before the study. By utilizing assessed data, the machine learning models aimed to predict whether or not a patient would continue to have PCS after four to five months. The random forest classifier achieved the highest performance with an 85\% accuracy and an area under the receiver operating characteristic curve (AUC) of 0.94. Factors found to be predictive of PCS outcome were Post-Traumatic Stress Disorder (PTSD), perceived injustice, self-rated prognosis, and symptom severity post-injury. The results of this study demonstrate that machine learning models can predict PCS outcomes with high accuracy. With further research, machine learning models may be implemented in healthcare settings to help patients with persistent PCS.

\end{abstract}
\smallskip


\smallskip
\smallskip
\smallskip

\section{Introduction}

Mild traumatic brain injuries, or concussions, are one of the most common brain injuries in the world, affecting as many as 6 per 1000 people a year~\cite{cassidy2004incidence}. While most concussion patients recover within several weeks, approximately 10\% of concussion patients continue to have symptoms post-injury~\cite{willer2006management}. Patients with persistent symptoms that last for more than several weeks are considered to have PCS. \par
Predicting whether or not a patient will continue to have PCS can be extremely important when providing PCS patients with proper care. First, since these models aim to predict PCS outcomes for patients already with prolonged symptoms, predictions can help physicians determine the length of further follow-up care necessary for the specific patient~\cite{bazarian2001predicting}. Similarly, predictions can be utilized to better determine when patients can return to normal activities. This could prove especially useful for patients who have taken significant time off from their jobs and need to know their PCS prognosis. \par
This paper aims to predict PCS outcome, or whether or not a patient will continue to have PCS after four to five months, using five different machine learning models: logistic regression, random forest classifiers, decision trees, k-nearest neighbors, and support vector machines.

\section{Related Studies}

Previous studies on PCS have shown that factors such as female gender, anxiety, symptom severity, and loss of consciousness are associated with PCS (\cite{ponsford2012predictors},\cite{ponsford2019factors},\cite{bazarian2001predicting},\cite{grubenhoff2014acute}). There have also been a number of studies in the past that have aimed to predict PCS using machine learning models. 

One such study assessed concussion patients in the emergency department to predict PCS at three months following the concussion. Measures of acute pain, postural stability, and neuropsychological functioning were used in a regression analysis model to classify patients. The regression model achieved 80\% sensitivity and 76\% specificity~\cite{sheedy2009emergency}. In another study, logistic regression and recursive partitioning were used to identify a high-risk group with a PCS risk of more than 90\% and a low-risk group with a PCS risk of less than 10\%~\cite{bazarian2001predicting}. Both studies focused on patients who had sustained a concussion within a 24-hour period of the study. For instance, the first study assessed participants at an average time of 13.9 hours after they had sustained a concussion~\cite{sheedy2009emergency}. 

While there have been some studies, such as the two mentioned above, that have aimed to predict PCS for patients who sustained concussions very recently, virtually no studies have been conducted on predicting outcomes in patients with persistent PCS. To address the lack of research on this specific group, the dataset chosen for this study consists of individuals with persistent PCS. More specifically, this study utilizes open data collected in a previous study, authored by Dr. Douglas P. Terry, regarding workplace and non-workplace concussions~\cite{terry2018workplace}. 

The open access article is distributed under the terms of the Attribution 4.0 International (Creative Commons By 4.0), which permits unrestricted use, distribution, and
reproduction in any medium, provided the author and source are credited. While the dataset included in Dr. Terry's paper is used in this study, this study has not been endorsed by him. The article can be accessed through a link that has been provided at the bottom of the paper. All participants were referred to specialized concussion clinics, where the study was conducted, because of their persistent symptoms post-injury. Thus, all patients were on atypical paths to recovery. During the initial assessment, participants were required to have sustained a concussion several weeks beforehand. More specifically, participants in the study sustained a concussion at an average of two to three months before the initial assessment. 

Finally, this study also aims to expand upon previous studies by utilizing a wider variety of machine learning algorithms and by implementing machine learning techniques such as grid search, feature selection, and stratified cross-validation to optimize model performance. 

\section{Methods}

\subsection{Participants and Procedures}
This study utilizes open data collected in a previous study regarding workplace and non-workplace concussions~\cite{terry2018workplace}. Patients who were referred to four different concussion clinics from 2015 to 2017 participated in the study, totaling 102 in number. At the initial assessment, participants had sustained a concussion within the past six months. During the initial assessment, injury characteristics such as loss of consciousness, initial British Columbia Postconcussion Symptom Inventory score (BC-PSI), self-rated prognosis, Injustice Experience Questionnaire (IEQ), and Post-Traumatic Stress Disorder Checklist (PCL-5) were measured. Four to five months after the initial assessment, participants were reassessed through telephone. Data collected during the second assessment was used to determine patient outcome (whether or not a patient continued to have PCS). Some participants could not be reached for the second assessment. 

During the second assessment, the new BC-PSI score was measured. As mentioned, this score was used to determine whether or not a patient continued to have PCS. Patients with a BC-PSI score of 15 or more, a very high score relative to healthy adults, were deemed to have clinically significant PCS~\cite{sullivan2011comparison}.

\subsection{Data Pre-processing}

All rows with null values were removed, reducing the number of rows from 102 to 81. Out of the 81 remaining participants, 53 had PCS during the second assessment, and 28 did not. After the missing values were removed, feature selection was conducted on the remaining data. 

Feature selection was conducted for several reasons. First, because there were 20 assessed variables at the initial assessment, feature selection was necessary to ensure the model was easier to interpret. (Note that while there were several variables assessed during the second assessment, none were used as input variables to train the models.) Furthermore, as the data had a large number of variables compared to the relatively small dataset, overfitting could occur if feature selection was not conducted. 

Thus, 20\%, or four input variables (self-rated prognosis, PCL-5, IEQ, and Initial BC-PSI), were used to train the machine learning models. Note that the ANOVA Test F-value was used as the feature selection method. 

\subsection{Variables}

\textbf{Self-rated prognosis:} During the initial assessment, participants were asked to self-rate their concussion prognosis. The possible answers were: “get better soon”, “get better slowly”, and “don’t know”. Previous studies have shown that self-rated prognosis is predictive of concussion outcome (\cite{cassidy2014population},\cite{kristman2016prognostic}).
\smallskip

\textbf{PCL-5:} The PCL questionnaire, also known as the Post-Traumatic Stress Disorder Checklist, measures PTSD symptoms. While questionnaire scores were collected for each individual question, the 20 questionnaire scores were combined to form one variable with a total score range of 0 to 80. (\cite{blevins2015posttraumatic},\cite{wortmann2016psychometric}).
\smallskip

\textbf{IEQ:} The Injustice Experience Questionnaire Total Score, ranging from 0 to 48, measures perceived injustice. Previous research has shown that the IEQ is valuable for measuring irreparability of injury and for measuring blame.~\cite{sullivan2008user}.

\smallskip

\textbf{BC-PSI:} BC-PSI was measured two times: the first, during the initial assessment, and the second, during the follow-up telephone call. The initial measure of the BC-PSI was part of the four variables used to train the models. The second assessment was used to determine the presence of clinically significant PCS. A participant with a BC-PSI score of 15 or more during the second assessment was considered to still have PCS. The BC-PSI score ranges from 0 to 52 and measures two aspects of PCS symptoms: frequency and severity~\cite{sullivan2011comparison}.

\subsection{Algorithms}

Logistic regression, random forest classifiers, decision tree models, k-nearest neighbors, and support vector machines were selected as the machine learning algorithms. Grid search was implemented on all models to find the optimal hyperparameters, with accuracy used to evaluate parameter settings. Stratified 5-fold cross-validation was also used during the training process. 

Likewise, when determining the receiver operating characteristic (ROC) curves and area under the ROC curves (AUC), stratified 5-fold cross-validation was used. Thus, for each model, five different ROC curves and five subsequent AUC values were computed, as well as a mean ROC curve and a subsequent AUC value. Note that all experiments were carried out on Google Colaboratory.

\section{Results}

The random forest classifier achieved the highest performance, with an accuracy of 0.85 and a standard deviation of 0.09. The mean AUC score of the model was 0.94, with a standard deviation of 0.10. The accuracy and AUC score of each model is shown below in Table \ref{table:1}.

\begin{table}[h!]
\centering
 \begin{tabular}{||c c c ||} 
 \hline
 Model & Accuracy (Standard Deviation) & AUC (Standard Deviation) \\ [0.5ex] 
 \hline\hline
 Logistic Regression & 0.78 (0.11) & 0.77 (0.15) \\ 
 Random Forest & 0.85 (0.09) & 0.94 (0.10) \\
 Decision Tree & 0.80 (0.09) & 0.73 (0.09) \\
 K-Nearest Neighbors & 0.81 (0.09) & 0.83 (0.16) \\
 Support Vector Machine & 0.80 (0.07) & 0.77 (0.05) \\ [1ex] 
 \hline

\end{tabular}

\smallskip
\smallskip
\smallskip

\caption{Model Performance}
\label{table:1}
\end{table}

\pagebreak
\begin{center} 

Receiver Operating Characteristic Curves
\end{center}

\begin{figure}[ht]
\begin{subfigure}{.5\textwidth}
  \centering
  % include first image
  \includegraphics[width=1.03\linewidth]{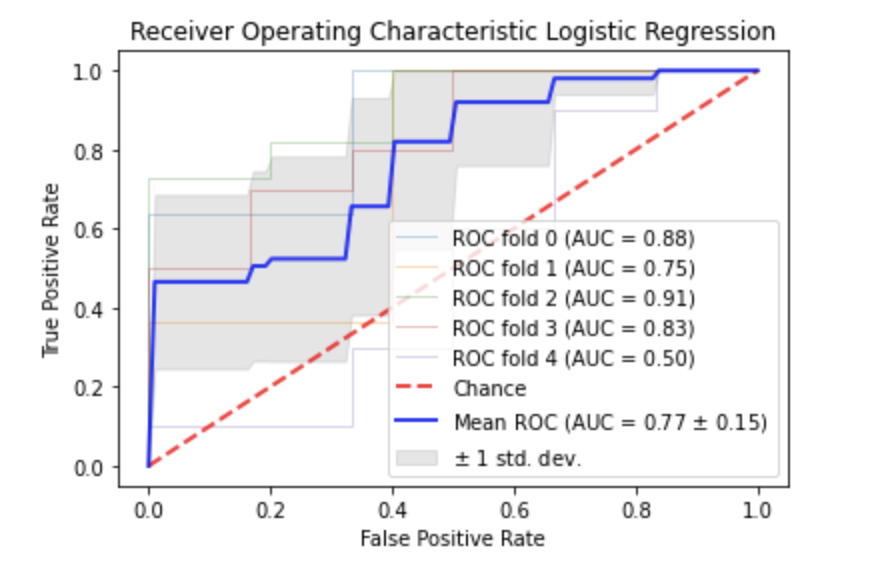}
  \caption{Logistic Regression}
\end{subfigure}
\begin{subfigure}{.5\textwidth}
  \centering
  % include second image
  \includegraphics[width=1.15\linewidth]{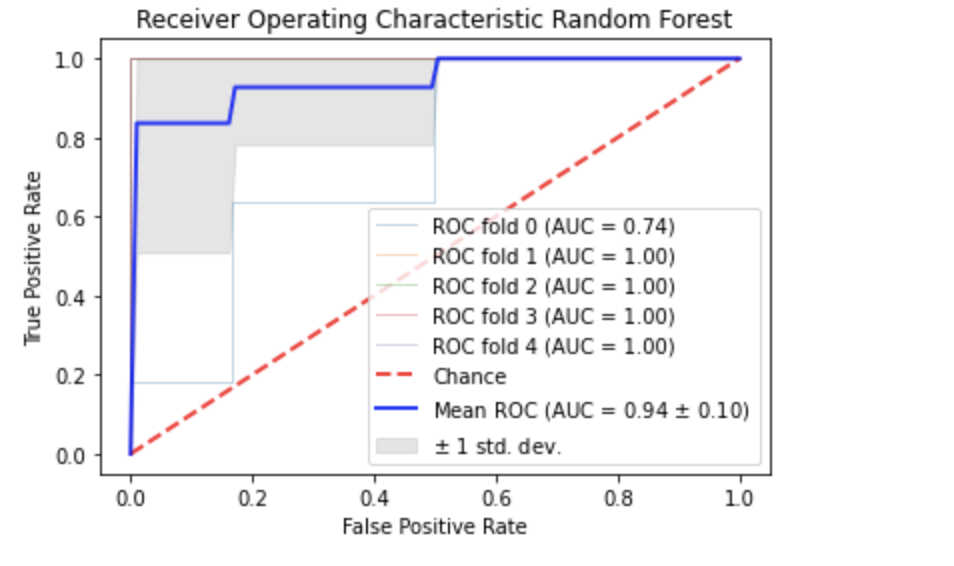}  
  \caption{Random Forest}
\end{subfigure}
\label{fig 1}
\end{figure}

\begin{figure}[ht]
\begin{subfigure}{.5\textwidth}
  \centering
  % include first image
  \includegraphics[width=1.1\linewidth]{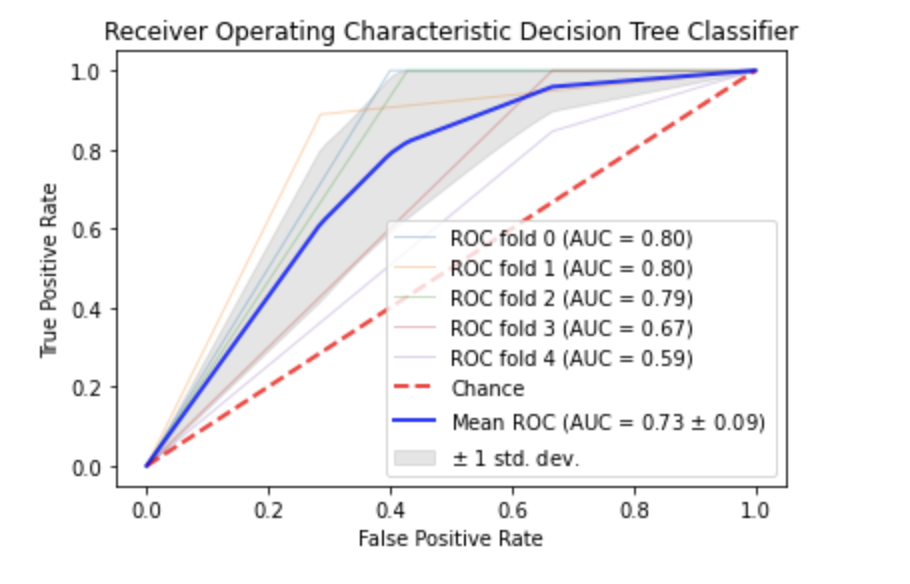}
  \caption{Decision Tree}
\end{subfigure}
\begin{subfigure}{.5\textwidth}
  \centering
  % include second image
  \includegraphics[width=1.05\linewidth]{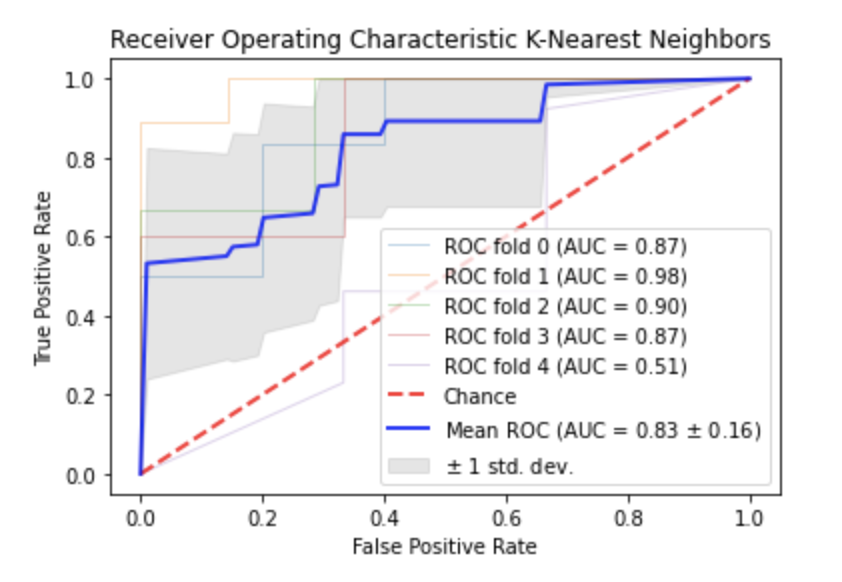}  
  \caption{K-Nearest Neighbors}
\end{subfigure}
\label{fig 2}
\end{figure}

\begin{figure}[H]
\begin{subfigure}{.5\textwidth}
  \centering
  % include first image
    \includegraphics[width=1.1\linewidth]{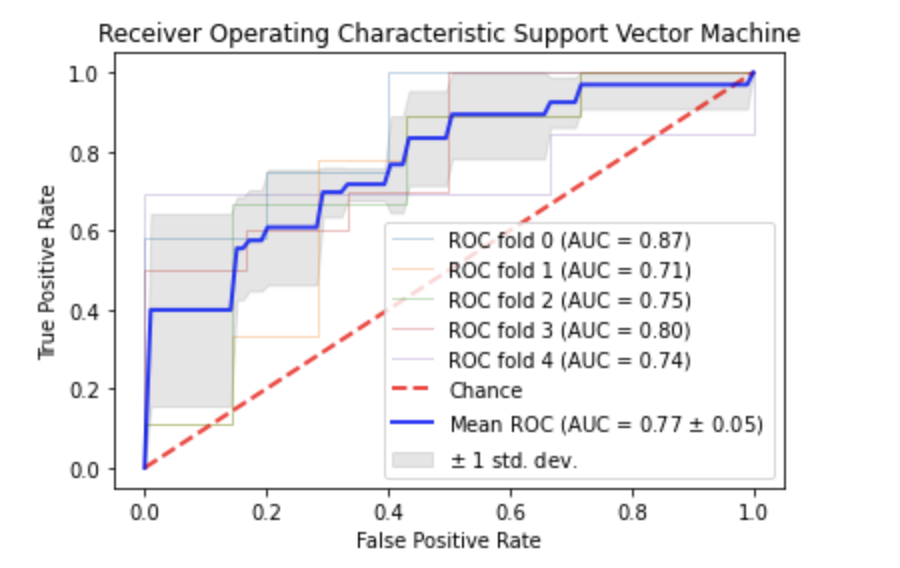}
      \caption{Support Vector Machine}
\end{subfigure}
\end{figure}

\section{Discussion}

All five models performed well in regards to model accuracy, with four models achieving accuracy scores above 80\%, and one model achieving a score of 78\%. In regards to AUC scores, one model achieved an outstanding score of 0.94, another model achieved an excellent score of 0.83, and the remaining three models achieved acceptable scores between 0.7 to 0.8~\cite{hosmer2013applied}. 

The standard deviations of the five models were relatively similar, but the AUC scores of the logistic regression model and the k-nearest neighbors had relatively high standard deviations, with standard deviations of 0.15 and 0.16 respectively. The high variability of the k-nearest neighbors model was expected, as k-nearest neighbors is generally a more flexible algorithm. However, the high variability of the logistic regression model was unexpected, as linear models tend to have low variability. In fact, the high variability of the logistic regression model suggests that the model was likely overfitted. Perhaps regularization methods could have been implemented along with the previous cross-validation and feature selection methods to prevent the model from overfitting. The standard deviations of the other models indicate how overfitting was avoided for those models.

Moving on, the results of the feature selection were somewhat unexpected, as the ANOVA Test F-value established four variables: self-rated prognosis, PCL-5, IEQ, and Initial BC-PSI. While previous papers have shown that these four variables are associated with PCS, other commonly associated variables such as loss of consciousness and female sex were not selected after feature selection (\cite{ponsford2019factors},\cite{bazarian2001predicting}). This may have been due to the particular dataset that was used in this study. Feature selection performed on other concussion datasets may have yielded different results.

While the results of this study are promising, further research needs to be conducted before models are implemented in a healthcare setting. First, models need to be trained on larger datasets to optimize performance. While methods such as cross-validation and feature selection were used to address some of the limitations of using small datasets, to further improve and validate the performance of the models, larger datasets are required. 

Similarly, datasets used may need to be more balanced. The dataset used in this study consists of 53 individuals with PCS during the second assessment and 28 individuals without PCS. 

Finally, a significant limitation of this study was the feature selection process. The ANOVA Test F-value was used to determine four input variables to train the models. While the four input variables that were used led to well-performing models, different feature selection methods could have produced even better results. While attempting to compute all possible subsets of input variables is virtually impossible due to the computational cost, methods such as feature selection by random search or other similar methods could be viable alternatives. 

\section{Conclusion}

Machine learning algorithms employed in this paper predicted outcomes for patients who had persistent PCS with high accuracy. The results of this paper are promising, and demonstrate that with further research, machine learning models may aid healthcare professionals in providing care for patients with persistent PCS.

\bibliographystyle{plainnat}
\bibliography{main.bib}

\smallskip
\smallskip
\smallskip
\smallskip
\smallskip

\emph{Email address}: minhongkim1@gmail.com

\smallskip
\smallskip
\smallskip
\smallskip
\smallskip
\smallskip

Link to paper authored by \emph{Dr. Douglas P. Terry:}

https://journals.plos.org/plosone/article?id=10.1371/journal.pone.0198128
\end{document}